\documentclass{webofc}
\usepackage[varg]{txfonts}   % Web of Conferences font

\usepackage[T1]{fontenc}
\usepackage[english]{babel}
\usepackage{amsmath}
\usepackage{float}
\usepackage{hyperref}
\usepackage{caption}
\usepackage{subcaption}

\newcommand{\thl}[1]{``#1''}
\newcommand{\nhl}[1]{#1}
\newcommand{\arxivlink}[2]{\href{https://arxiv.org/abs/#1}{\texttt{arXiv:#1}}}
\newcommand{\cdslink}[1]{\href{http://cds.cern.ch/record/#1}{\texttt{cds:#1}}}
\newcommand{\Sec}{Section~}

\newcommand{\Fig}{Fig.~}

\newcommand{\law}{Law}
\newcommand{\luigi}{Luigi}

\begin{document}
\title{End-to-End Analysis Automation over Distributed Resources with Luigi Analysis Workflows}

\author{\firstname{Marcel} \lastname{Rieger}\inst{1}\fnsep\thanks{\email{marcel.rieger@cern.ch}}}

\institute{Institute of Experimental Physics, Hamburg University, Luruper Chaussee 149, 22761 Hamburg}

\abstract{%
    In particle physics, workflow management systems are primarily used as tailored solutions in dedicated areas such as Monte Carlo production.
    However, physicists performing data analyses are usually required to steer their individual, complex workflows manually, frequently involving job submission in several stages and interaction with distributed storage systems by hand.
    This process is not only time-consuming and error-prone, but also leads to undocumented relations between particular workloads, rendering the steering of an analysis a serious challenge.
    This article presents the Luigi Analysis Workflow (\law) Python package which is based on the open-source pipelining tool \luigi, originally developed by Spotify.
    It establishes a generic design pattern for analyses of arbitrary scale and complexity, and shifts the focus from executing to defining the analysis logic. 
    \law\ provides the building blocks to seamlessly integrate with interchangeable remote resources without, however, limiting itself to a specific choice of infrastructure.
    In particular, it introduces the concept of complete separation between analysis algorithms on the one hand, and run locations, storage locations, and software environments on the other hand.
    To cope with the sophisticated demands of end-to-end HEP analyses, \law\ supports job execution on WLCG infrastructure (ARC, gLite, CMS-CRAB) as well as on local computing clusters (HTCondor, Slurm, LSF), remote file access via various protocols using the Grid File Access Library (GFAL2), and an environment sandboxing mechanism with support for sub-shells and virtual environments, as well as Docker and Singularity containers.
    Moreover, the novel approach ultimately aims for analysis preservation out-of-the-box.
    \law\ is developed open-source and independent of any experiment or the language of executed code, and its user-base increased steadily over the past years.
}

\maketitle

\section{Introduction}
\label{sec:intro}

\begin{figure}[h!tbp]
        \begin{center}
            \includegraphics[width=0.50\textwidth]{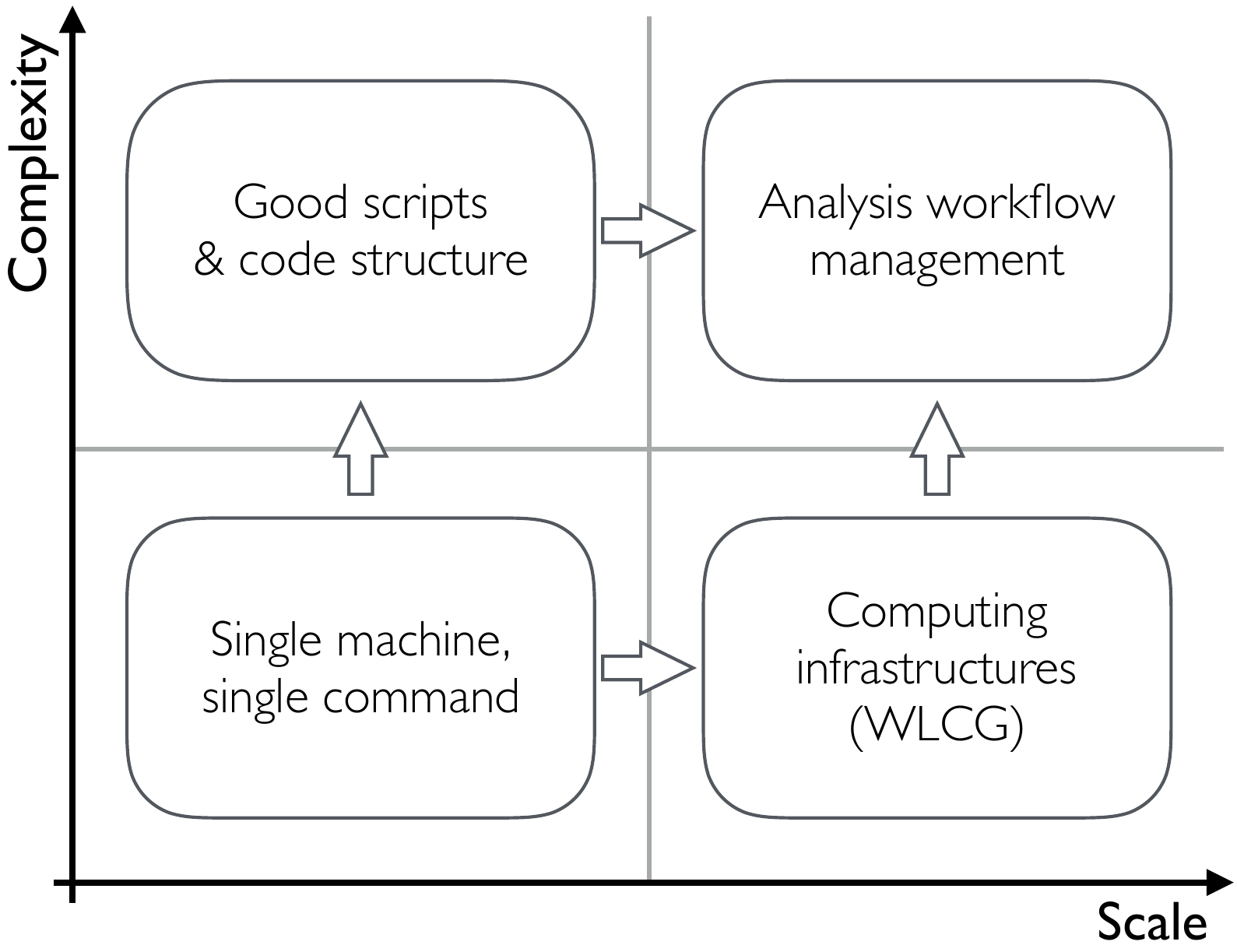}
            \caption{
                Scale and complexity as specification measures for physics analyses and their impact on the choice of structural conception.
            }
        \label{fig:scalecomplexity}
        \end{center}
    \end{figure}

    The management of scientific workflows presents a complex challenge in today's physics working environments.
    High-level physics analyses usually consist of a considerable amount of logically separated workloads.
    In general, the interface between these workloads does not rely on an event-by-event data flow.
    The workloads to accomplish a specific research question form a loose collection of inhomogeneous procedures, encoded in executable files such as Shell and Python scripts, and are executed manually.
    Hereby, their execution order is dictated by the dependencies between them in terms of the existence of results of one or more previous procedures.
    Beyond a certain scale and complexity, i.e., degree of granularity and inhomogeneity of workloads, manual steering of analysis workflows can be time-consuming, prone to errors, and potentially leads to undocumented relations between workloads.
    The interplay between scale and complexity, and their impact on the analysis conception are indicated in \Fig\ref{fig:scalecomplexity}.
    
    A novel design pattern for physics analyses conception and automation that copes with the challenges of inhomogeneous workload definition and risks due to manual steering is presented.
    It is based on the pipelining package \luigi\ \cite{luigi:2018} due to its simple yet scalable and extensible design, providing guidance on structuring arbitrary workloads.
    To meet the sophisticated demands of performance-intensive analyses, the developed software, called \textit{Luigi Analysis Workflows} (\law), allows to seamlessly integrate remote resources as used in, yet not exclusively, high-energy physics.
    Conceptually, \law\ is independent of any experiment, specific use case, programming language, and data format.
    In typical high-energy physics nomenclature, it does not qualify as a \thl{framework}, but it rather defines a set of permissive guidelines and tools to follow a design pattern for distributed analyses.
    The eventual goal of the software is to provide a versatile working environment for robust analyses on scalable and interchangeable resources, while encouraging analysis preservation.

    Specially tailored solutions already exist for a small number of particle physics applications such as the comprehensive event simulation campaigns performed centrally by the CMS and ATLAS experiments at the LHC\ \cite{cms:2005:tdr_computing,atlas:2005:tdr_computing}.
    However, requirements for performing end-to-end physics analyses can differ greatly.
    Particular workloads may require the results of multiple previously executed workloads and produce more than one output themselves.
    These dependencies form a directed acyclic graph (DAG), which can take arbitrary shapes potentially even depending on dynamic run-time conditions.
    In addition, the shape of the DAG is not necessarily known a-priori and also might be subject to continuous development cycles.
    Furthermore, whereas central experiment workflows often rely on dedicated infrastructure deployments, analyses must rather incorporate existing resources and maintain the ability to adapt to short-term changes.

    This article is structured as follows.
    First, design guidelines encouraged by the developed package are described in \Sec\ref{sec:guidelines}.
    In \Sec\ref{sec:luigi}, the central components of the \luigi\ package are presented.
    The technological key concepts of the developed \law\ software are explained in \Sec\ref{sec:law}.
    They comprise remote execution, remote storage, and environment sandboxing.
    Resulting prospects for analysis preservation are summarized subsequently, before presenting conclusions in \Sec\ref{sec:concl}.

\section{Design Guidelines}
\label{sec:guidelines}

\begin{figure}[h!tbp]
        \centering
        \includegraphics[width=0.69\textwidth]{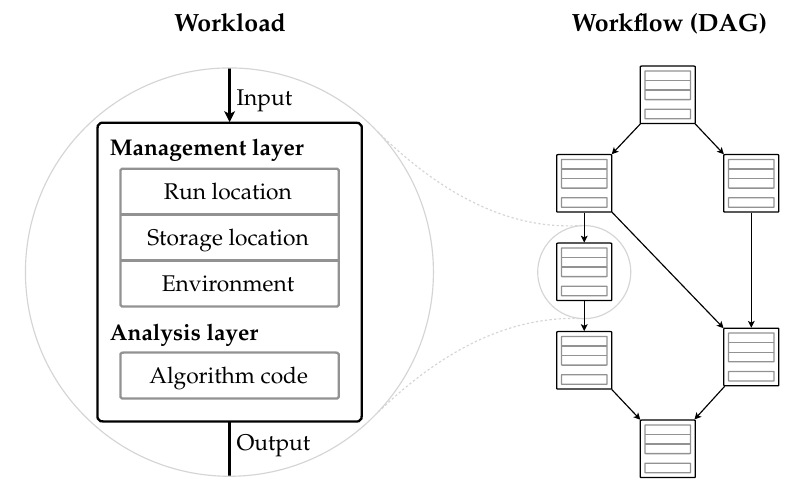}
        \caption{
            Illustration of a workflow composed of multiple workloads, forming a directed acyclic graph (DAG).
            The management layer defines and handles run location, storage location, and environment.
            The analysis layer comprises the actual algorithm to run and is fully decoupled from the management layer, allowing for interchangeability of resources.
        }
        \label{fig:workflow_generic}
    \end{figure}

    The key operating principles of the presented software and the associated analysis design pattern are based on four guidelines, which are discussed in the following paragraphs.
    A supporting illustration is presented in \Fig\ref{fig:workflow_generic}.
    \begin{description}
        \item[1. Encapsulation]
        Logically separated steps of an analysis are encapsulated in workloads.
        They can produce outputs, require the completion of certain other workloads, and therefore consider their outputs as potential inputs.
        Workloads and the dependencies between them compose a workflow that is visualizable as a DAG.
        The graph structure is scalable, versatile, and leverages automatic dependency resolution which can be exploited by execution systems to process entire workflows using a single command invocation.

        \item[2. Factorization]
        A workload is factorized into four domains that are organized in two layers.
        The \thl{management} layer handles the run location of a workload, defines the location where output data is stored, and steers the software environment during execution.
        The \thl{analysis} layer contains the actual algorithm code to run.
        It is fully decoupled from the management layer, i.e., algorithms remain unaffected by any decisions about run and storage locations.
        Dynamic behavior of both layers can be achieved by including parameters to the workload.
        Their values might be configured at execution-time or depend on run-time conditions.

        \item[3. Interchangeability]
        Implementations of domains of the management layer are interchangeable.
        For instance, the change of the storage location neither affects the analysis algorithm, nor decisions about its run location or software environment.
        Evidently, exceptions occur in cases where special hardware requirements (e.g. GPUs for machine learning applications), or strict dependencies between run and storage locations exist (e.g. data access on remote computing elements behind firewalls is often restricted to dedicated storage systems).
        % provides a high degree of flexibility and
        The interchangeability of resources prevents hard-coded dependence on certain infrastructure.
        Decisions on all three domains can be made at the execution-time, allowing, for instance, switching between local and remote resources for testing purposes.

        \item[4. Universality]
        The realization of algorithms in the analysis layer and the type, quantity, and content of their results are universal.
        There is no limitation on the programming language or implementation details of algorithms, as long as they are executable by means of sub-processing.
        Before and after the actual processing, measures can be taken to ensure the compliance with the factorization (2) and interchangeability (3) paradigms.
        Furthermore, the form of analysis results is entirely unrestricted, i.e., any kind of stateful information can serve as a workload output.
        This includes files with arbitrary data formats, database entries, or even specific file content, e.g. contained in machine-parsable job reports.
        Universality causes the distinction from the concept of \thl{frameworks} in the sense of typical particle physics applications.
    \end{description}

    Compliance with the four guidelines above entails several accomplishments.
    Relations between workloads are encoded as dependencies and become an integral part of an analysis.
    Besides the evident advantages for automation, this also leads to notable benefits for the purpose of documentation.
    Furthermore, the DAG defined by a workflow can be visualized with appropriate tools and, in environments such as provided by standard web browsers, can even serve as a basis for status monitoring during execution.
    Also, beyond a certain scale and complexity, manual steering of physics analyses can be a challenging task.
    Especially in working groups where analyses are fully or partially operated by multiple individuals, a common interface between workloads has significant advantages.
    Final and intermediate results are both reproducible and reusable between similar analyses, effectively lowering the degree of unnecessary redundancy.

    Technically, the guidelines introduced above are realizable using extensible concepts provided by the \luigi\ package, which is introduced in the following.

\section{\luigi}
\label{sec:luigi}

    \luigi\ is a Python software package that provides a scalable design pattern for structuring large and complex workflows\ \cite{luigi:2018}.
    Initially developed at Spotify, it became a community-driven, open-source project and is successfully deployed in both commercial and scientific applications.
    The execution model follows a \texttt{make}-like approach as it only computes what is really necessary in order to produce the output of a requested workload \cite{make:2016}.
    Its features entail automatic failure handling, command-line interfaces, web visualization, and file system abstractions.
    The following paragraphs introduce the fundamental building blocks.

    In \luigi, an arbitrary workload, i.e., the elementary unit in an overarching workflow, is described as a \thl{task}.
    The purpose of a task is to produce a customizable set of outputs, denoted as \thl{targets}.
    While targets usually represent local or remote files, they can, in principle, describe any type of stateful resource.
    The sole core functionality of a target is to check and report its own existence.
    Therefore, a task is considered \thl{complete} when all its output targets exist.
    Moreover, to alter the default behavior of a workload, tasks can expose and implement mandatory and optional \thl{parameters}.
    Significant task parameters, i.e., parameters that influence the content of output targets, should be encoded in the respective target locations, e.g. via distinctively integrating them into file paths.
    This injective mapping is encouraged to prevent ambiguities during checks of a task's completeness condition.
    Finally, tasks can \thl{require} one or multiple other tasks to define a coherent workflow, where outputs of required tasks become \thl{inputs} to the current task.
    As tasks can implement requirements depending on values of particular parameters, workflows can depend on highly dynamic behavior and adapt to a large variety of use cases.

\begin{figure}[b!]
        \begin{minipage}[b]{0.44\textwidth}
            \includegraphics[width=1.0\textwidth]{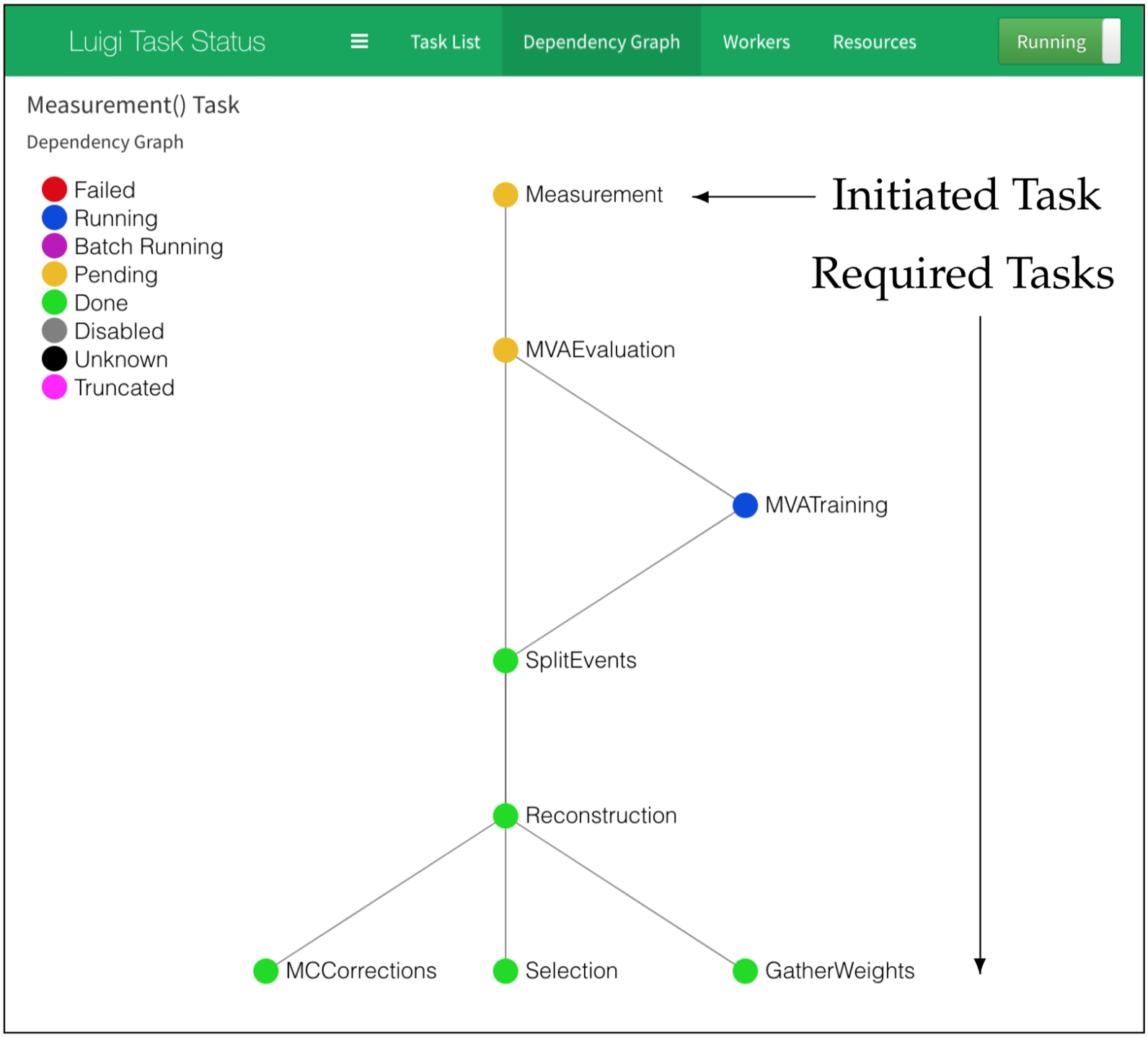}
        \end{minipage}\hfill
        \begin{minipage}[b]{0.50\textwidth}
            \caption{
                Screenshot of the \luigi\ web visualizer application, showing a hypothetical analysis workflow consisting of an initiated task and its recursive requirements in a directed acyclic graph (DAG).
                Edges visualize direct dependencies and node colors denote task statuses.
            }
            \label{fig:luigi_overview}
        \end{minipage}
    \end{figure}

    A workflow that consists of interdependent tasks can be described as a directed acyclic graph (DAG), and visualized upon execution using a web application provided by \luigi.
    An example is shown in \Fig\ref{fig:luigi_overview} for a hypothetical analysis.
    The execution of a workflow is initiated by running a particular task.
    \luigi\ recursively evaluates the completeness condition of all required tasks to infer the shape, i.e., nodes and edges of the underlying DAG.
    Subsequently, tasks are scheduled according to their position in the DAG with the number of parallel processed tasks as well as additional priority rules being configurable.
    As a result, \luigi's stateful execution system is deterministic and resource-effective since it only (re)processes what is really necessary.

    For physics applications, workflow management tools such as \luigi\ can give rise to considerable advantages.
    Large-scale analyses with complex dependency structures can be modeled via tasks based on a design pattern consisting of only a few building blocks, while keeping overhead on a manageable level.
    Besides the separation of workloads and a consistent parameterization approach, all dependencies are expressed as part of the analysis with direct benefits for error prevention, documentation, and collaboration.
    At the same time, neither constraints on the programming language nor on the format of data that is exchanged between tasks is introduced.

    Further features include automatic failure handling, command-line interface generation per task, task templates that support map-reduce jobs and file system abstractions (such as \nhl{Hadoop} and \nhl{HDFS}\ \cite{shvachko:2010}).

\section{\law}
\label{sec:law}

    The \law\ software package extends \luigi\ to include remote resources available in the context of high-energy physics research.
    It is designed as a layer on top of \luigi\ rather than as a replacement, without changing any of its core functionalities.
    % It is written in Python, which is a common choice in scientific applications owing to its high acceptance, support for diverse programming patterns such as object orientation and scripting, and the vast amount of available libraries.
    % Workloads are described by parameterizable \thl{tasks} that may require other tasks, define output \thl{targets}, and separately designate the algorithm code to run.
    % Outputs of required tasks denote inputs, and a task is considered complete when all its output targets exist.
    The guidelines introduced in \Sec\ref{sec:guidelines} are realized through a set of technological key concepts, namely remote execution of tasks, remote storage of output targets, and software environment sandboxing.
    In the subsequent paragraphs, they are discussed in detail, followed by a discussion on implications for analysis preservation.

    \paragraph{Remote Execution}
    Physics analyses investigating rare processes in high-energy particle collisions typically examine a considerable amount of data to infer statistically significant results.
    This implies that a large quantity of computing tasks must be executed, which most likely exceed the capacity of local host systems in terms of processing units, memory consumption, and storage space.
    Consequently, remote task execution on distributed batch systems is imperative.
    Following the principle of factorization as introduced above, analysis algorithm code should remain unaffected by any decisions on run locations.

    The developed software realizes remote execution capabilities through \thl{mixin-based} inheritance.
    Tasks can inherit from one ore more classes that provide functionality for submitting computing jobs to certain remote batch systems. %as well as for retrieving job status information and results.
    % In contrast to inheritance models that denote \thl{is-a} relationships, mixin-based inheritance is considered as a technique to adapt common functionality while avoiding code redundancy.
    Thus, the configuration of run locations is rather declarative and, in particular, does not interfere with other factorized workload domains.
    In case of multiple available run locations provided by different mixin-classes, the actual implementation for a particular batch system can be selected at execution time using task parameters, which fulfills the demand for interchangeability.
    At the time of writing, the \nhl{HTCondor}\ \cite{tannenbaum:2002,thain:2005}, \nhl{LSF}\ \cite{lumb:2004}, \nhl{ARC}\ \cite{ellert:2007}, \nhl{gLite}\ \cite{andreetto:2008}, \nhl{Slurm}\ \cite{yoo:2003} batch job interfaces, as well as the CMS Remote Analysis Builder (CRAB)\ \cite{mascheroni:2015} system are supported.

    % TODO: remove?
    % Another strict policy of the remote execution system is customizability.
    % Deployed instances of the same batch system, such as \nhl{HTCondor}, can behave fairly different regarding submission requirements and overall supported features, depending on how they are configured by administrators.
    % Therefore, the implemented execution system neither assumes a particular setup nor does it attempt to extrapolate specifications, but rather provides transparent access to all job configuration options.
    % Furthermore, various features are included which are crucial for efficient operation on batch systems.
    % Among others, they include job error detection, automatic resubmission, early stopping criteria, and the possibility to connect to common monitoring interfaces.

    \paragraph{Remote Storage}
    \label{sec:law:remote_storage}
    The large amount of data to be processed inevitably causes the need for distributed data storage systems.
    Requirements on the storage capacity per analysis in the order of $\sim 10 - 100$\,TB are not unusual.
    Especially when relying on large-scale remote execution systems such as the WLCG\ \cite{cms:2005:tdr_grid}, the use of high-throughput storage is mandatory.
    % TODO: remove 1
    % Furthermore, storage locations should be accessible from within run locations in order to avoid manual copying of files and permanently occupying redundant disk space.

    The concept of output targets represents a suitable abstraction for handling files on interchangeable remote resources.
    While multiple realizations of targets are conceivable in general, the following paragraphs focus solely on physical files.
    Independent of whether targets denote local or remote files, they strictly implement the same common interface, which defines both low- and high-level file operations.
    Algorithms in the analysis layer can rely on the equivalence of operations, fully omitting distinction of cases to achieve interchangeability of storage locations.
    Additionally, the approach can be used to enable load distribution over several different locations.
    The respective behavior is configurable on target- and task-level.

    The default interface for remote targets in \law\ is based on the open-source Grid File Access Library (GFAL2) \cite{ayllon:2017}, which enables file access through various protocols such as provided by \nhl{XRootD}, \nhl{dCache}, \nhl{SRM}, \nhl{GridFTP}, and \nhl{Amazon S3}.
    Moreover, the developed interface provides essential features including transfer validation, automatic retries for robustness against connection disruptions, and optional local caching.
    The latter is of particular importance for the optimization of network utilization in a variety of use cases such as repeated local tests.
    Both cache synchronization and invalidation are fully integrated into all remote target operations.

    \paragraph{Environment Sandboxing}
    \label{sec:law:environment_sandboxing}
    % TODO: remove 1
    Another important aspect of the factorization paradigm is the management of software and environments.
    Different workloads within the same analysis workflow may depend on distinct software setups that are not necessarily compatible. % with each other.
    Examples are workloads that utilize experiment-specific software that is often subject to extensive validation procedures and therefore requires long-term stability.
    This is opposed to machine learning algorithms built on top of third-party libraries, which may crucially improve in performance when updated regularly.
    For this purpose, an environment \thl{sandboxing} mechanism is integrated into \law.

    A sandbox describes a collection of environment configurations, ranging from plain sets of variables over to different operating systems.
    Current implementations support subshells with custom initialization files, Python virtual environments, CMS software (CMSSW) environments\ \cite{cms:2005:tdr_computing}, and containers realized through \nhl{Docker} \cite{boettiger:2015} or \nhl{Singularity} \cite{kurtzer:2017}.
    Tasks can declare the demand for a particular sandbox and, optionally, provide further logic for negotiating with the current environment to determine if either compatibility is sufficient or sandboxing is necessary.
    In the latter case, the developed mechanism embeds the invocation of the algorithm code within the specified sandbox.
    This approach preserves the encapsulation paradigm by considering environment dependencies as part of the management layer of workloads and thus, enables the execution of inhomogeneous workflows with a single command.

    Another benefit of sandboxing is related to long-term stability and portability.
    Most host systems and computing clusters are subject to constant changes owing to maintenance and security measures.
    For workloads with critical software dependencies, sandboxing provides a method for ensuring reproducibility of results over time.
    Therefore, it exhibits a key technique for the preservation of analyses, which is discussed in the following paragraph.

    \paragraph{Analysis Preservation}
    \label{sec:law:analysis_preservation}
    The preservation of analyses is a key ingredient for sustainability of experimental physics research.
    Long-term storage of data, analysis algorithms, and related external information is required to ensure the repetition of measurements with reproducible results for the purposes of documentation, education, and reinterpretation.
    Especially reinterpretation campaigns represent an important use case as they could utilize preserved analyses to test hypotheses postulated by newly developed theories, exploiting the amount of recorded data and developed algorithms to this date.
    Inter-experimental collaborations have formed to advance this common effort \cite{dphep:2012}.
    In practice, however, the process proves to be challenging owing to the fact that analyses need to be adapted manually to fulfill certain preservation conditions.
    In contrast, analysis workflow management systems with integrated preservation capabilities could offer appealing benefits.
    On the basis of the four guidelines introduced in \Sec\ref{sec:guidelines}, a decisive set of preservation conditions can be derived:
    \begin{enumerate}
      \item
      The analysis layers of workloads containing algorithm implementations must not be changed.
      Moreover, operations relying on random number generators should use fixed seeds.

      \item
      Software and other relevant environment configurations must be retained such that repeated execution of the analysis code leads to identical results.
      % External libraries might be updated when complying with prescriptions of semantic versioning.

      \item
      All initial and external files must either be retained or referenced to copies on dedicated data preservation systems.

      \item
      Access to remote infrastructures providing parallel execution and storage systems must be configurable for analyses that require a considerable amount of resources.

      \item
      The execution of analysis workflows must be fully automated and\,/\,or sufficiently documented in order to guarantee operability by other researchers at any time.
    \end{enumerate}
    Analyses fulfilling all of the above aspects possess low barriers for enabling preservation.

    \law\ intrinsically satisfies conditions 1, 4 and 5 by adhering to the principles of workload domain factorization and interchangeability of run and storage locations.
    The retention of environment configurations (condition 2) can be achieved by means of the sandboxing mechanism.
    Here, containerization methods such as provided by \nhl{Docker} \cite{boettiger:2015} or \nhl{Singularity} \cite{kurtzer:2017} are particularly suitable.
    For high-energy physics experiments, the preservation of simulated and recorded collision events (condition 3) requires sophisticated measures.
    Integration with LHC open-data initiatives \cite{cowton:2015} is mandatory and therefore encouraged by \law.

\section{Conclusions}
\label{sec:concl}

    The presented guidelines and tools for generic analyses conception constitute a novel approach for coping with the increasing demands of modern high-energy physics data analysis.
    The \luigi\ package is a viable solution to address the complexity of structuring and executing workloads in a \texttt{make}-like fashion.
    The developed \law\ package adds scalability in the scope of high-energy physics infrastructure in a non-intrusive way by interfacing common job submission systems and remote data storage.
    In addition, a customizable sandboxing mechanism ensures the integrity of software and computing environments, and thus the reproducibility of physics results.

    As the described approach does not introduce constraints on software or data structures, it is considered a toolbox providing an \thl{analysis design pattern} rather than a \thl{framework}.
    The resulting workflows preserve all information about entangled analysis workloads with benefits for automation, error prevention, and documentation.
    In a broader context, the presented project provides the means to extend the concept of collaboration beyond the sharing of code.
    Eventually, the resulting increase of transparency and reproducibility paves the way for analysis preservation.

    The source code is publicly available under BSD license \cite{rieger:2018}.


\end{document}